# Pre-pairing and polarization of doped charge carriers in unconventional superconductors

*Lixin Zhang*, Mengyan Nie†, Yan Cheng†, and Ran Zhao†*

School of Physics, Nankai University, Tianjin 300071, People's Republic of China

**ABSTRACT:** Based on first-principles calculations, we show in hole-doped $La_{2-x}Sr_xCuO_4$, the Sr dopants prefer to form dimers, rather than be isolated as generally believed. Same trend applies to other superconductors such as Na-doped $Ca_2CuO_2Cl_2$, and electron type $Nd_{2-x}Ce_xCuO_4$. The dimer structure makes the two doped holes (electrons) bound up and form a singlet which is also polarized. Another type of dopant in cuprates is oxygen which behaves similarly. So, we propose that, for all oxide high $T_C$ superconductors, the holes (electrons) are pre-paired (Z-pairs) during the preparation of the superconductor. The spontaneously formed Z-pairs are conceptually different from not only the Cooper pairs, but also the Zhang-Rice singlets. The attraction force in a Z-pair originates basically from the strong antiferromagnetic exchange energy. We also found that the on-site Coulomb repulsion (U) is no longer a constant due to doping. Around a doping site, U can reduce its value discontinuously, which plays a key role in stabilizing the Z-pairs against isolated holes (electrons). A Z-pair is a p-wave like localized state which breaks the symmetry of the $CuO_2$ plain and attracts each other along the direction perpendicular to the polarization. The Z-pairs may be popular in all of the unconventional superconductors and the periodic assembly can result in the emergence of coherence states and superconductivity.

* E-mail: lxzhang@nankai.edu.cn

†These authors contribute equally.

## INTRODUCTION

This year marks the 40$^{th}$ anniversary of discovery of high $T_C$ superconductivity in a cuprate by Bednorz and Müller [1] which had triggered intensive studies ever in the history of Physics. However, till now the theoretical understanding of the materials is still very poor. This phenomenon could have already made history if not for the whole physics world but at least for the field of condensed matters. What is wrong and what mistake had we made in the past four decades? It is time to reflect seriously in order to move forward quickly.

What is unique for the high $T_C$ superconductors compared with other materials studied in the condensed matter history? It is the extremely high density of defects as of ~10%. This indicates that the number of defects is ~$10^7$ times larger than that in a semiconductor which is of ~$10^{15}$. The number of the defects is just one order of magnitude less than the host atoms which indicates that the interaction between the defects is extremely complicated and strong, and can't be simply neglected. We all know as condensed matter physicist that interaction between electrons (Coulomb repulsion) in a solid is the key to the vast amount of different properties of the solid materials. Similarly, the interaction between the defects could be the origin of the vast amount of unconventional properties of the high $T_C$ materials. Unfortunately, in the past 40 years it has never been taken seriously in the theory development. Many authors still rely on concepts developed for the Fermi liquids and conventional superconductors due to lack of appropriate theories. This could be just the reason why we had made a history of four decades of searching in the dark.

Is there any experimental evidence that the interaction between defects in a cuprate superconductor is strong and affects the superconductivity? Actually, there are so many. For example, STM and ARPES experiments [2-6] both report that with doping, highly localized in-gap states start to emerge within the so-called charge transfer gap (CTG) of the un-doped Mott insulator. These states originate or shifted from the upper Hubbard band (UHB). Most strikingly, with increasing of the defect density, the shift of the in-gap states both from UHB to CTG and to the deeper levels in the CTG gap continues. A theoretical study [7] claims that a defect may act as a repulsive potential that can move the in-gap states in the Hubbard bands. However, the details of the interaction between defects and its relationship to superconductivity has never been studied.

In Fig. 1 we show a schematic of the electronic structure of the Mott insulator and the evolution of the states with doping. We see that the in-gap states from UHB have a sister state, the hole states from lower Hubbard band (LHB). This indicates, the doping affects not only the in-gap states which locate high above $E_F$, but also the hole states, which locate very close to $E_F$, and closely relates to superconductivity.

In this study, based on density functional theory, using a series of computational material science methods, we studied the atomic structure of the unconventional superconductors, focusing on the defect related structural and electronic properties under proper density. What we found is that the interactions between defects can be attractive and the defects tend to organize themselves into various new orders. With the ordering of the dopants, the localized holes (electrons) originate from them order accordingly. The finding is really astonishing and contradictory to what people usually believed. It might be just the simple truth that we omitted in the past 40 years.

## DETAILS AND DISCUSSION

In this part, we will show the details of our calculational results, focusing on a varying U distribution in the doped cuprates. In the present, it is still very hard to determine the exact values of U. However, based on the trend observed from experiment, we can determine the formation energy of a dopant dimer and its dependent on the U change. We find that a dimer configuration for the dopants is much favored in doped cuprates. The dimer structure is associated with two localized holes on the $CuO_2$ plain, which are paired spontaneously. The paired holes break the symmetry of the $CuO_2$ plain and are polarized along either of the Cu-O bond directions. From our experience, it seems that only the polarized hole (electron) pairs relate to the high $T_C$ superconductivity, not the single holes, or the so called Zhang-Rice singlets [8].

In the whole paper, we try to focus on the common features of all unconventional superconductors, not the difference, in order to have a simple, clear and unified picture of the mechanism of such superconductivities.

**The origin of the reduced U in doped cuprates**: In the parent compounds, it is well known that the on-site Coulomb repulsion (U) is the key to understand the insulating nature of the compounds. In Fig. 1, we show the related energy levels schematically. We can see the band gap results from U. It is clear that if we reduce the U locally, the band gap can reduce locally either. Not only the in-gap states, but also the hole states will be affected by the reduction of U. They are sister states and closely related, for example, their localizations might be similar. This could be the reason why some authors in their experimental reports do not distinguish them properly.

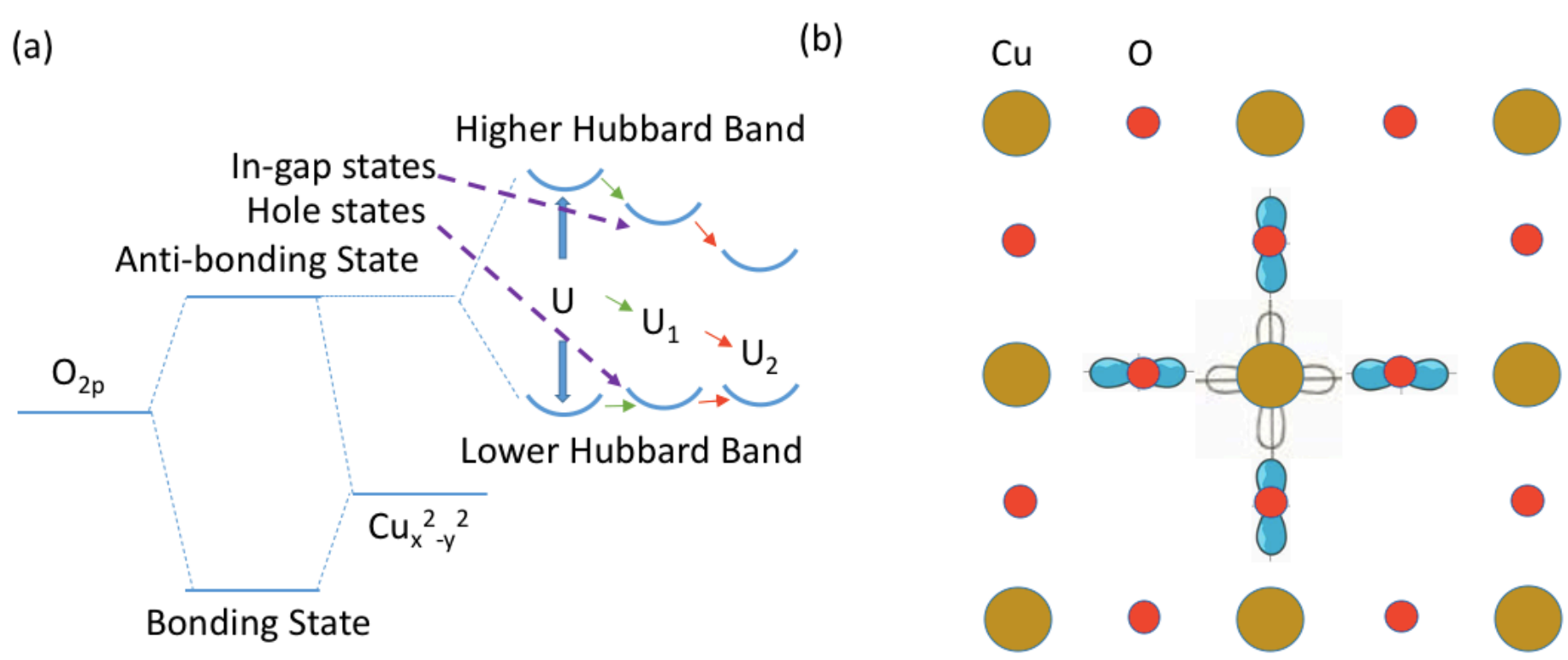

Fig.1 Schematic plots (a) for the levels of the Cu and O orbitals, their bonding and antibonding states, and the Hubbard bands due to U on a $CuO_2$ plain. The continuous evolution of the Hubbard bands to localized in-gap states and hole states is the result of doping with higher and higher densities. (b) for the $d_{x^2-y^2}$ and $p_x$, $p_y$ orbitals in real space on the $CuO_2$ plain.

Many experiments have clearly shown that the in-gap states are evolved from upper Hubbard band with doping [2, and references there in]. These states are highly localized around the defect site and the level of the states are sensitive to the geometry of the site, i.e., the number of defects in the neighborhood and their distance and orientation which is complex. However, a general trend can be concluded that the higher the density of the defects, the lower the levels of the in-gap states.

We can reasonably suppose that the in-gap states result from the reduction of U on the nearest neighbor Cu sites. For $Ca_2CuO_2Cl_2$， our calculations show that U = 8eV corresponds to a bandgap of 2.06 eV. If it is reduced to 7eV (6eV) only on the nearest neighbor Cu sites for a Ca vacancy, a peak of in-gap states is at 1.74 eV (1.46 eV). So, the observation of the in-gap states with doping is a direct result of the local reduction of U on Cu atoms around the defect. What we are interested is whether such reduction could affect the calculation of the formation energies of the defects.

**Dopant dimerization in cuprate high $T_C$ superconductors**: First we will check the structural property of Sr substitutionals in $La_{2-x}Sr_xCuO_4$. We calculated a 4x4x1 supercell with two Sr substitutionals in one LaO layer. In one configuration, the two Sr substitutionals are next to each other to form a Sr dimer as shown in Fig. 2a. In another configuration, two Sr atoms are located as far as possible as two monomers. The total energy difference (ΔE) for the two configurations are shown in Fig. 2b. For the first configuration, we used U and U-$\Delta U_2$ on different Cu sites as shown in Fig. 2a. For the second configuration, we used U and U-$\Delta U_1$ (on nearest neighbor Cu) accordingly. The steps for $\Delta U_1$ and $\Delta U_2$ are both 0.2 eV. Fig.

2b show the ΔE for a series of ($\Delta U_1$, $\Delta U_2$). The green dashed line corresponds to ΔE=0 and the white dashed line is for $\Delta U_1 = \Delta U_2$. The upper left parts in Figs. 2c-2d satisfy $\Delta U_1 < \Delta U_2$, which corresponds to real situations as previously discussed. It is clear that for all the reasonable combinations of ($\Delta U_1$, $\Delta U_2$), $\Delta E < 0$, which means that the dimer configuration is always favorable than the monomers one.

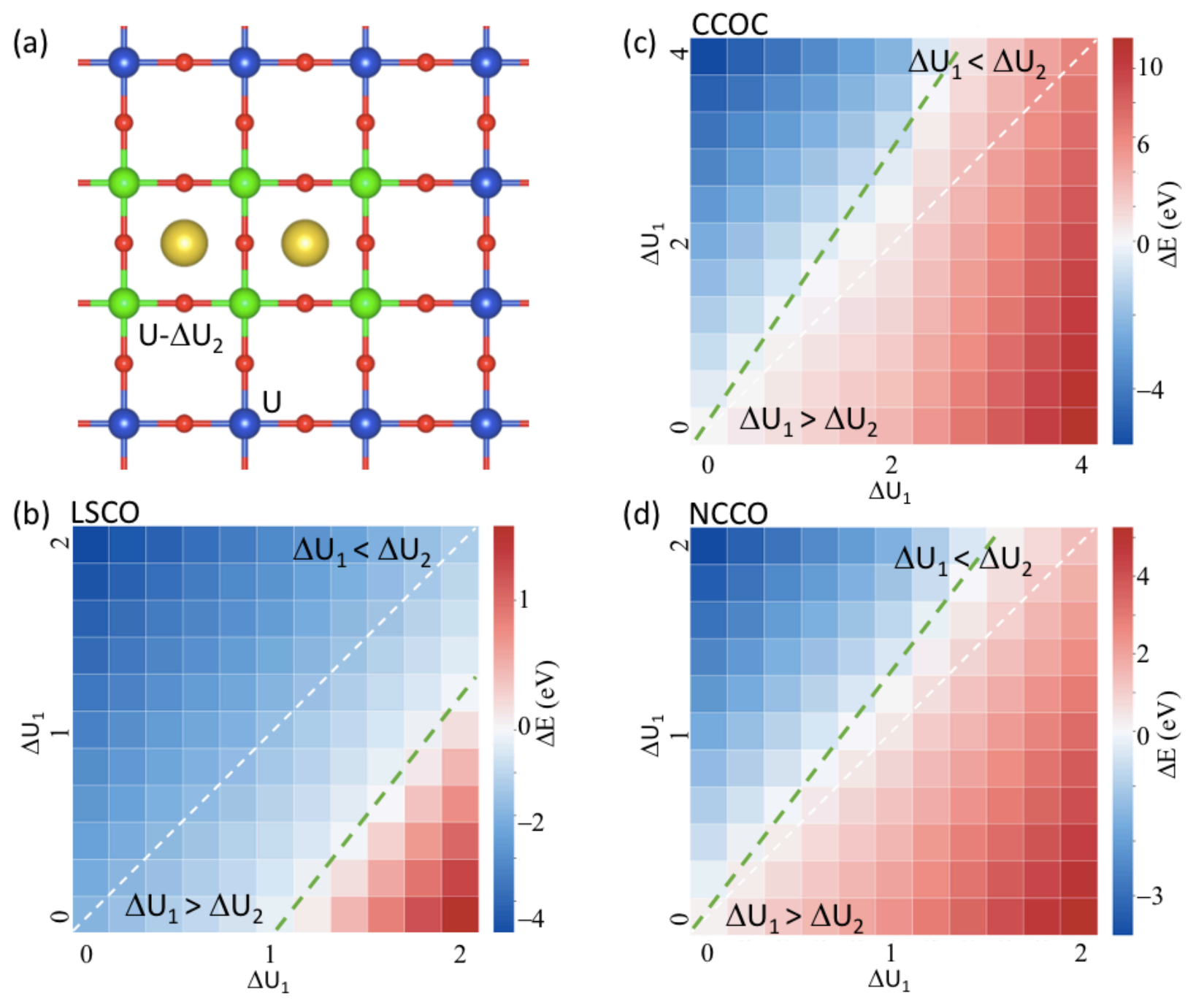

Fig.2 The total energy difference (ΔE, in eV) between a dimer configuration (as shown in (a)) and the isolated monomers configurations (not shown) for the hole dopants in (b) $La_{2-x}Sr_xCuO_4$, (c) Na-doped $Ca_2CuO_2Cl_2$, and (d) electron type $Nd_{2-x}Ce_xCuO_4$ with various combinations of ($\Delta U_1$, $\Delta U_2$). In (a) it is shown that the U is reduced to U-$\Delta U_2$ on nearest neighbor Cu atoms. In the monomers configuration, it is reduced to U-$\Delta U_1$ on the nearest Cu neighbors. The white dashed lines in (b) to (d) divide the region into two parts respectively. The upper left parts are for $\Delta U_1 < \Delta U_2$ corresponds to real situations. The green dashed lines run cross most of the ΔE=0 regions.

Fig. 2c is for Na-doped $Ca_2CuO_2Cl_2$. In the region $\Delta U_1 < \Delta U_2$, $\Delta E > 0$ only takes place for combinations with high $\Delta U_1$ which is unlikely because for isolated defects, the in-gap states are usually shallow and $\Delta U_1$ is small. Fig. 2d is for electron type $Nd_{2-x}Ce_xCuO_4$. The situation is similar to Na-doped $Ca_2CuO_2Cl_2$.

In short summary, we see that, for all the three cuprate superconductors, where the dopants are one valence difference from the substituted atoms, the dopant atoms prefer to form dimers, rather than isolated monomers.

Another important factor for the stability of a defect configuration is its entropy in the lattice. The entropy depends on density of the defect. At low density, where entropy is large, $\Delta E < 0$ can't guarantee the free energy of dimer configuration being favored. This is to say that the dopant atoms still prefer to be isolated at finite temperature. The situation may correspond to the AFM phase for underdoped cuprates in the phase diagram of high $T_C$ superconductors. At high density, especially when x~10%, the doping sites is only one magnitude smaller than the host atoms, the entropy is small as $\sim k_B T$, which can be neglected in the free energy calculation. Now $\Delta E < 0$ can guarantee a minimum free energy for the dimer configuration.

We also checked the stability of clusters of three dopant atoms for $La_{2-x}Sr_xCuO_4$. The most favored structure has the configuration of one dimer and one monomer. Then, the formation energies for each Sr substitution are 5.63-($\mu_{Sr}$-$\mu_{La}$), 5.11-($\mu_{Sr}$-$\mu_{La}$), and 5.47-($\mu_{Sr}$-$\mu_{La}$) ($\Delta U_1 = \Delta U_2 = 0$), where $\mu_{Sr}$ ($\mu_{La}$) is the chemical potential of Sr (La), for monomer, dimer, three-atom structures respectively. It is clear that to form Sr dimers is the mostly favored.

**The concept of Z-pair and its polarization**: The above discussion indicates that in a doped high $T_C$ superconductor, the dopant atoms can be pre-paired during material preparation. Then what is the impact to the electronic structure of cuprates and the relation to their superconductivity? Let's examine the electronic structure of a dopant dimer first. Each atom in the dimer will donate one hole (electron) to the $CuO_2$ plain. That is, the two holes (electrons) are donated to the same local region and a bound state is formed. The two holes should have opposite spins and they form a singlet. This singlet is totally different from the Zhang-Rice singlet which is composed of one electron and one hole. It is something like a Cooper pair but is different in the way how the two charge carriers are paired. Let's name it Z-pair. A Z-pair contains two holes or two electrons, with different spins that form a singlet, bound together by dimerization of dopant atoms or by one dopant like oxygen (will be discussed later). Since each hole (electron) is s-wave like due to the symmetry of the $CuO_2$ plain as shown in Fig. 3a, a Z-pair is composed of two s-wave states with a fixed distance $a_0$ (the lattice constant of the $CuO_2$ plain), like a p-orbital, or half of the lobes of a d-orbital, as shown in Figs. 3c and 3d. Unlike Cooper pairs, who can intermix in the space freely, a Z-pair can only locate along one of the Cu-O bond directions. So, it is polarized on the $CuO_2$ plain. Two Z-pairs with opposite polarizations can form a d-wave like state effectively.

Consider a Ca vacancy in $Ca_2CuO_2Cl_2$ [2]. Although two holes are doped to the $CuO_2$ plain from one Ca vacancy, the two holes are both s-wave and the combination of the two holes is still a s-wave, as shown in Fig. 3b. The two holes also form a pair but it is totally different from the polarized ones in Figs. 3c and 3d. Consider the fact that we have not seen any report about the superconductivity of Ca-vacancy doped $Ca_2CuO_2Cl_2$, we propose that the existence of holes (electrons) does not guarantee the emergence of superconductivity. On the other hand, polarization and ordering of the Z-pairs may play the key role.

We need talk more about oxygen as a dopant in high $T_C$ superconductors. It is well known that oxygen doping usually results in higher $T_C$. The highest $T_C$ is found in cuprates by oxygen doping. However, the structure of the oxygen doped cuprate is usually inhomogeneous. With increasing doping, the structure is getting more and more homogeneous, with the ordering of oxygen dopants in a fractal type structure [9]. Let's take a single oxygen in the Cu-O chain in orthorhombic $YBa_2Cu_3O_{6+\delta}$ as an example. Two 2p orbitals on such oxygen are interested. One bounds with the Cu atoms along the chain, one provides the holes to the $CuO_2$ plain with p-orbitals perpendicular to the Cu-O chain, as shown in Fig. 3d. Thus, the two holes also form a polarized pair on the $CuO_2$ plain.

As shown in Fig. 3a, one hole keeps the C4 symmetry of the $CuO_2$ plain. For the Ca vacancy case, the C4 symmetry is kept either as shown in Fig. 3b. Fig. 3c shows the case for a dopant dimer which can be regarded as the superposition of two holes with a distance of $a_0$. Clearly the C4 symmetry is broken by the polarized Z-pair. Fig. 3d shows a case of the oxygen on a Cu-O chain in $YBa_2Cu_3O_{6+\delta}$. The C4 symmetry is broken either. So, we can see that the polarized hole pairs break the C4 symmetry of the $CuO_2$ plain and the interaction between the Z-pairs will be anisotropic and polarization dependent.

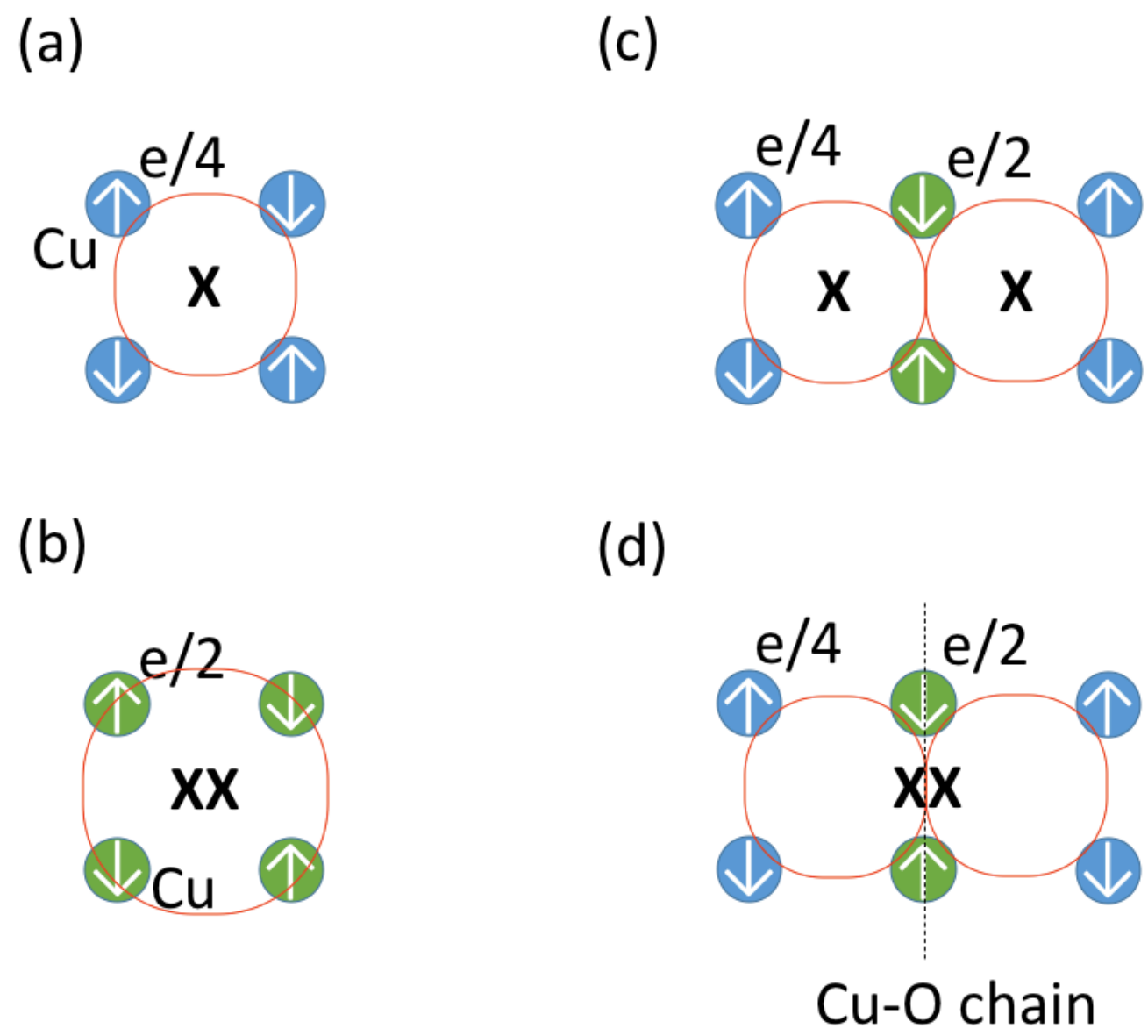

Fig. 3 Schematics for the polarizations of holes in a high Tc superconductor. **X** indicates the position of a one-hole dopant, and **XX** of a two holes dopant. Numbers of nominal charges (e/4 and e/2) on each Cu atoms are shown. (a) an isolated hole, (b) two holes from a Ca vacancy, (c) a polarized hole pair by a dopant dimer, (d) a polarized hole pair by a two-hole dopant like oxygen. The charges distributed on the next nearest neighbors and beyond are neglected. The red circles show schematically the shapes of the wavefunctions of one hole (electron) and two holes. In (c) and (d) the two holes with a fixed distance $a_0$ form a p-wave like state. Two perpendicular p-wave states form a d-wave like state.

Another interesting thing is that as in $La_{2-x}Sr_xCuO_4$, the doped holes on $CuO_2$ plain can be from two adjacent LaO layers. So, we have two types of interactions for the Z-pairs. One is between the Z-pairs from the same LaO layer, one is between Z-pairs from different LaO layers. This could be the reason why $La_{2-x}Sr_xCuO_4$ has relatively lower $T_C$. For $YBa_2Cu_3O_{6+\delta}$, however, one oxygen on the Cu-O chain may donate the Z-pair to two adjacent $CuO_2$ plains. Thus, the adjacent $CuO_2$ plains are correlated and this could be just the reason why the $T_C$ of $YBa_2Cu_3O_{6+\delta}$ is higher and the highest $T_C$ is found in cuprates with three $CuO_2$ plains in one unit cell.

**Mechanism for the Z-pair formation:** The energy of an antiferromagnetic lattice can be described by a simple Hamiltonian $H = -\sum'_{l,l'} J\overrightarrow{S_l} \cdot \overrightarrow{S_{l'}}$, where $J < 0$, and $l, l'$ for nearest neighbors only. This means that the exchange energy $J$ keeps the system stabilized. To form one hole, four nearest neighbor bonds are broken and the energy of the system will be increased by - $4J > 0$ (with $E_F$=0). For two isolated holes, the energy increased is - $4J$ x2 = - $8J$. On the other hand, to form a hole pair as shown in Fig. 4a, 6+1 nearest neighbor bonds are broken and the energy of the system is increased by - $7J < -8J$. Thus, a Z-pair is energetically favored over two isolated holes. This is also true for other 2D antiferromagnetically coupled insulators.

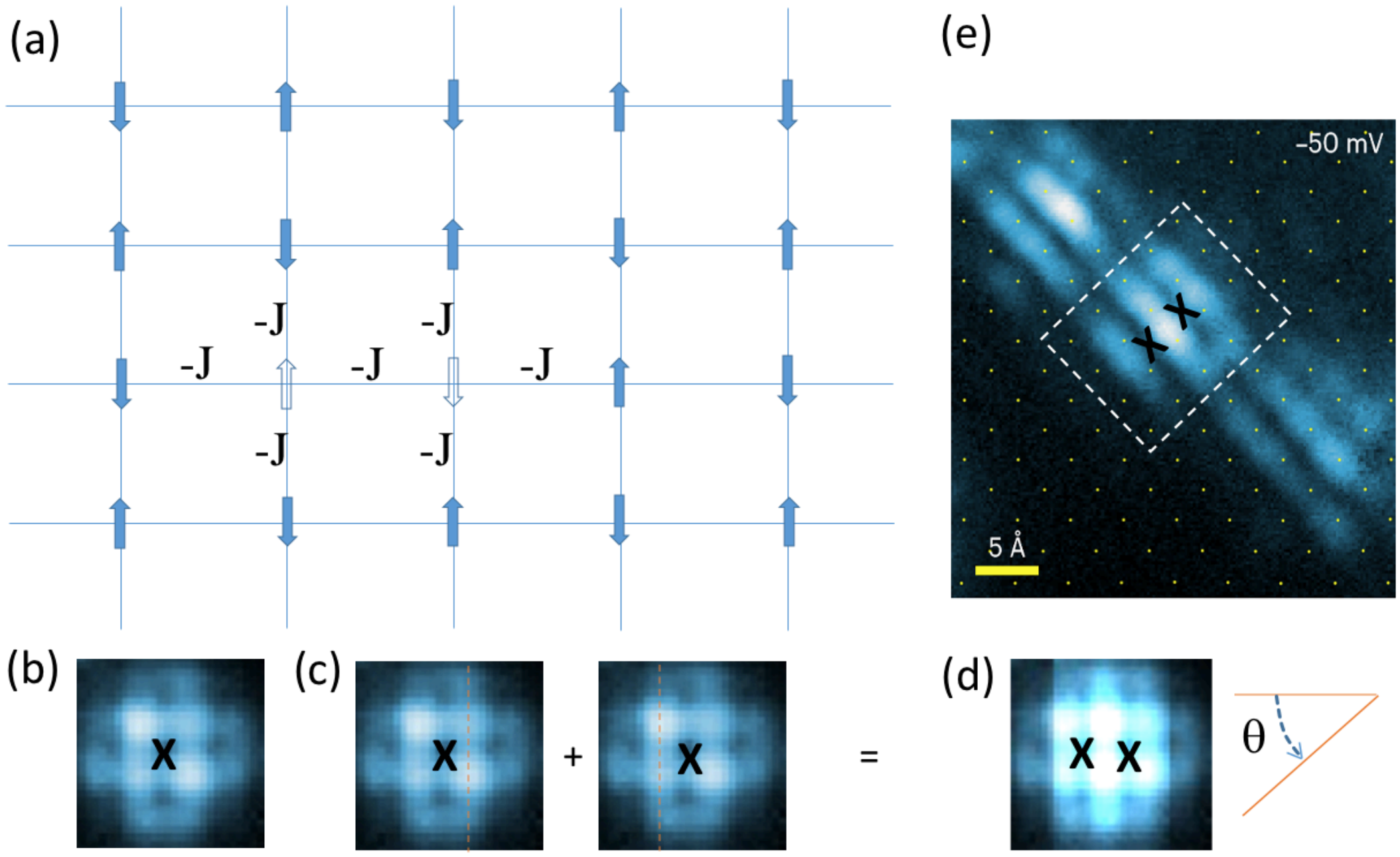

Fig. 4 (a) Schematic plot for the increase of energy of two doped holes on $CuO_2$ plain. (b) is taken from Fig. 1d in reference [10] where a magenta dot is replaced by "X" to indicate the dopant location. (c) showing the two lines to align with for the merge of the two copies of (b). (d) is the result of the merge and an angle to rotate for the plot in order to compare with the squared area in (e) which is taken from Fig. 2c in reference [10].

**Experimental evidence for the Z-pairs**: Using the concepts developed here, it seems that a lot of puzzles can be explained easily for the unconventional superconductors, at least qualitatively. Here we show a few of them. First, in Fig. 4b, a STM image is take from reference [10]. It is more reasonable to regard it as a Na dopant, or one isolated hole rather than a Zhang-Rice singlet as the authors claim. We simply merge two copies of the image by aligning with the dotted lines as shown in Fig. 4c and have Fig. 4d. By rotating an angle of θ it can be directly compared with the squared area in Fig. 4e, which is take from the same reference. We can see that they are quite similar, especially the symmetries. This indicates that the plaquette in the chequerboard image in reference [10] corresponds to a Na dimer, i.e., a polarized Z-pair. Then the chequerboard structure is a 2D array of Na dimers, i.e., a 2D array of polarized Z-pairs. We notice that the authors of the paper have mentioned that each plaquette in the chequerboard image has about two holes. Our understanding is that, the two holes are pre-paired and the pairs are uniform and polarized. Within a stripe in the image, the polarizations of the plaquettes are the same as discussed in the following.

Due to the polarization of a Z-pair, the C4 symmetry of the $CuO_2$ plain is broken and the interactions between Z-pairs would be anisotropic. Our calculations of $La_{2-x}Sr_xCuO_4$ show that, Z-pairs align along the Cu-O direction that is perpendicular to the Sr-dimer is favored by 1.55 eV over parallel to the dimer. The optimal distance is actually $4a_0$, agreeable with many experimental reports, which might be a characteristic length for the Z-pairs on $CuO_2$ plain.

At the beginning of the doping, the dopant atoms may distribute as monomers randomly, due to the entropy effect. With increasing of the impurity density, dimers start to appear, with equal opportunities to take two possible polarizations. With further doping, the dimers will attract other dimers with the same polarization to form a periodic chain of Z-pairs along one of the Cu-O bond direction, which corresponds to the stripes in the STM images. This kind of ordering of dopants have long been observed by experiments [9] to highly correlate with the $T_C$ of the superconductor.

Another evidence is from reference [11]. In this report, the authors claim that their RIXS experiments have shown even-ordered phonon excitations, which they speculate could originate from localized Cooper pairs. From our understanding, the localized Cooper pairs are just the polarized Z-pairs due to the pre-pairing of Sr dopants in the sample.

Because the distance between the Z-pairs is $4a_0$, the optimal density of the holes is 1x2/(4x4) = 1/8 = 12.5% for a full layer of Z-pairs, which is very close to the densities for optimal $T_C$ . Because not all holes or hole pairs are polarized Z-pairs, the optimal density from experiment usually larger than 12.5%. For superconductors with oxygen dopants, the optimal density could be very close to this value [12] because, all holes are in the form of polarized Z-pairs.

Many experiments [12] such as Knight shift and specific heat measurements, point out to singlet formation in doped cuprates as the origin of the pseudogap phenomenon. Likely, the concept of Z-pairs developed here can explain easily how the singlet is formed and why the temperature scale reaches high above $T_C$ for the pseudogap.

A direct way to confirm the existence of dopant dimers is by XRD. The dimers have short range order along Cu-O bond directions of the $CuO_2$ plain.

With the doped holes (electrons) pre-paired in singlets, we are facing a two fluids system. One is the RVB states [13], the other is the Z-pairs. Electron pairs flow between the two fluids by pair scattering. With this picture in mind, we can probably find a state with the lowest free energy and $T_C$. The key is that whether we can understand how the Z-pairs interact with each other on one $CuO_2$ plain and between the $CuO_2$ plains.

We also studied twisted bilayer graphene, another unconventional superconductor. We found theoretically that the holes are favored to be doped in pairs to same unit of the Moirè lattice. Another ongoing study in our group is the one-atomic layer FeSe on a substrate. Similar result is emerging, that is, the charge carriers prefer to pair in real space, rather than in k-space as in BCS theory [14].

For a quite long time, we understand that the doped cuprates are no longer metals we are familiar with and the superconductivity is unconventional. However, till now, many researchers are still stick to concepts of Fermi liquid and BCS picture of electrons pairing. Our study and a recent report [15] both point out to a real space pairing for unconventional superconductors. The Fermi surface is hard to discern. Isolated holes (electrons) may destroy superconductivity. The reciprocal space is no longer convenient. It is just time to set up a totally new theory for all unconventional superconductors.

**CONCLUSIONS**

To conclude, we show two findings in this paper. One is that the dopants in cuprates may have strong attractive interactions and form dimers. The result is that the holes (electrons) are pre-paired instead of being isolated and moving freely. The pre-paired charge carriers are polarized and can attract each other to assemble periodically along specific directions. This could be also a general phenomenon for all unconventional superconductors. The other is that the on-site Coulomb repulsion U in such materials is no longer a constant value with doping. Our first finding is based on such a reduction. High $T_C$ superconductivity could be the result of the ordering of the polarized pre-paired charge carriers (Z-pairs).

**ACKNOWLEDGMENT**

The work was supported partly by the NSF of China with grant No. 11574157. Many former undergraduate and graduate students Dazhuang Deng, Yuli Han, and Wenhao Wu had done related calculations and we thank their contributions to the work. Haoyang Zhang plotted parts of Fig.2 and Fig. 4, which is deeply appreciated.

## COMPUTATIONAL DETAILS

The calculations were performed within the framework of density functional theory as implemented in the Vienna *Ab initio* Simulation Package (VASP) [16, 17]. The Kohn-Sham wave functions were expanded in a plane wave basis set with a cut-off energy of 500 eV. The projector-augmented wave (PAW) method and Perdew-Burke-Ernzerhof (PBE+U) potential for the exchange-correlation functional were used [18]. For $La_2CuO_4$, $Ca_2CuO_2Cl_2$, and $Nd_2CuO_4$, U's of 6.5 eV, 8eV, and 3.2 eV on Cu were applied. Supercells of 4×4×1 were used for the defect formation energy calculations. The optimization is not stopped until the maximum force acting on each atom is less than 0.02 eV/Å. The Monhkorst–Pack *k*-point mesh is 2×2×2. Dipole corrections are applied. The formation energy of substitutional defect is calculated by:

$$E_f = (E_s + n_s\mu_s) - (E_b + n_D\mu_D) \qquad (1),$$

where $E_s$ and $E_b$ are the energies of the doped and bulk supercell, $n_s$ and $n_D$ are the numbers of substituted atoms and dopant atoms, $\mu_s$ and $\mu_D$ are their chemical potentials respectively.

## REFERENCES

[1] J. G. Bednorz, & K. A. Müller, Possible high Tc superconductivity in the Ba–La–Cu–O system. Z. Phys. B64, 189–193 (1986).
[2] H. Li, S. Ye, J. Zhao, C. Jin, and Y. Wang, Sci. Bull. 66, 1395(2021).
[3] C. Ye, P. Cai, R. Yu, X. Zhou, W. Ruan, Q. Liu, C. Jin, and Y. Wang, Nat. Commun. 4, 1365(2013).
[4] P. Cai, W. Ruan, Y. Peng, C. Ye, X. Li, Z. Hao, X. Zhou, D. H. Lee, and Y. Wang, Nat. Phys. 12, 1047(2016).
[5] Y. Wang, Y. He, K. Wohlfeld, et al., Commun. Phys. 3 Article number: 210 (2020).
[6] Z. Chen, Y. Wang, S. N. Rebec, T. Jia, M. Hashimoto, D. Lu, B. Moritz, R. G. Moore, T. P. Devereaux, and Z. X. Shen, Science 373, 1235(2021).
[7] C.-P. He, et al., Chinese Phys. Lett. 39, 057401(2022).
[8] F. C. Zhang and T. M. Rice, Phys. Rev. B 37, 3759(R) (1988).
[9] M. Fratini, N. Poccia, A. Ricci, et al., Scale-free structural organization of oxygen interstitials in La2CuO4+y. Nature 466, 841–844 (2010).
[10] S. Ye, J. Zhao, Z. Yao, et al., Visualization of the Zhang–Rice singlet, electronic molecules and Cooper pair formation in a cuprate superconductor. Nat. Phys. 22, 1401–1408 (2026).
[11] Ke-Jun Xu, Paulina Majchrzak, Nathan Giles-Donovan, Sangheon Kim, Jaewon Choi, Jun Okamoto, Ganesha Channagowdra, Hsiao-Yu Huang, Di-Jing Huang, Seiki Komiya, Takao Sasagawa, Shuichi Wakimoto, Wei-Sheng Lee, Zhi-Xun Shen, Thomas P. Devereaux, Dung-Hai Lee, and Robert J. Birgeneau, arXiv:2609.08076v1 [cond-mat.str-el]
[12] P. A. Lee, N. Nagaosa, and X.-G. Wen, Doping a Mott insulator: physics of high-temperature superconductivity. Rev. Mod. Phys. 78, 17–85 (2006).
[13] P. W. Anderson, et al., The physics behind high-temperature superconducting cuprates: the plain vanilla version of RVB. J. Phys. Condens. Matter 16, R755–R769 (2004).
[14] J. Bardeen, L. N. Cooper, and J. R. Schrieffer, Phys. Rev. 108, 1175 (1957).
[15] Zhen Zhu, Yudi Huang, Julian May-Mann, Kaiming Liu, Zheyu Wu, Shanta R. Saha, Johnpierre Paglione, Alexander G. Eaton, Andrej Cabala, Michal Vališka, Eduardo Fradkin, Vidya Madhavan, Evidence of intertwined pair density and charge density wave orders in UTe2. Proceedings of the National Academy of Sciences, 123, 29(2026).
[16] G. Kresse, and J. Hafner. Phys. Rev. B 47, 558 (1993).
[17] G. Kresse, and J. Furthmuüller. Phys. Rev. B 54, 11169 (1996).
[18] J. P. Perdew, K. Burke, and M. Ernzerhof. Phys. Rev. Lett. 77, 3865 (1996).